\documentclass{aa}

\usepackage{amsmath}
\usepackage{amssymb}
\usepackage{graphicx}

\newcommand{\bfr}{{\boldsymbol{r}}}
\newcommand{\bfv}{{\boldsymbol{v}}}

\newcommand{\txd}{{\text{d}}}

\renewcommand{\leq}{\leqslant}
\renewcommand{\geq}{\geqslant}

\DeclareMathOperator{\arccotg}{arccotg}
\DeclareMathOperator{\arcsecans}{arcsec}
\DeclareMathOperator{\arcsech}{arcsech}
\DeclareMathOperator{\fracoperator}{frac}
\DeclareMathOperator{\intoperator}{int}

\begin{document}

\title{%
The Hernquist model revisited:\\
completely analytical anisotropic dynamical models}

\titlerunning{The Hernquist model revisited}

\author{Maarten Baes\thanks{Postdoctoral Fellow of
the Fund for Scientific Research, Flanders, Belgium
(F.W.O.-Vlaanderen)} \and Herwig Dejonghe}

\authorrunning{M. Baes \& H. Dejonghe}

\offprints{M. Baes}

\institute{Sterrenkundig Observatorium, Universiteit Gent,
Krijgslaan 281-S9, B-9000 Gent, Belgium\\
\email{maarten.baes@rug.ac.be, herwig.dejonghe@rug.ac.be}}

\date{Submitted 13 June 2002; Accepted 16 July 2002}

\abstract{Simple analytical models, such as the Hernquist model,
are very useful tools to investigate the dynamical structure of
galaxies. Unfortunately, most of the analytical distribution
functions are either isotropic or of the Osipkov-Merritt type, and
hence basically one-dimensional. We present three different
families of anisotropic distribution functions that
self-consistently generate the Hernquist potential-density pair.
These families have constant, increasing and decreasing anisotropy
profiles respectively, and can hence represent a wide variety of
orbital structures. For all of the models presented, the
distribution function and the velocity dispersions can be written
in terms of elementary functions. These models are ideal tools for
a wide range of applications, in particular to generate the
initial conditions for $N$-body or Monte Carlo simulations.
}

\maketitle

\section{Introduction}

From a stellar dynamical point of view, the most complete
description of a stellar system is the distribution function
$F(\bfr,\bfv)$, which gives the probability density for the stars
in phase space. In this paper, we will concentrate on the problem
of constructing anisotropic equilibrium distribution functions
that self-consistently generate a given spherical mass density
profile $\rho(r)$. In the assumption of spherical symmetry, the
mass density of stellar system can easily be derived from the
observed surface brightness profile, at least if we assume that
the mass-to-light ratio is constant and that dust attenuation is
negligible. And as the surface brightness of a galaxy (or bulge or
cluster) is fairly cheap and straightforward to observe, compared
to other dynamical observables which require expensive
spectroscopy, the problem we will deal with is relevant and
important.

The first step in the construction of self-consistent models is
the calculation of the gravitational potential $\psi(r)$, which
can immediately be determined through Poisson's equation. The
second step, the actual construction of the distribution function,
is less straightforward. Basic stellar dynamics theory (see e.g.\
Binney \& Tremaine 1987) learns that steady-state distribution
functions for spherical systems can generally be written as a
function of binding energy and angular momentum. We hence have to
determine a distribution function $F({\cal{E}},L)$, such that the
zeroth order moment of this distribution function equals the
density, i.e.\ we have to solve the integral equation
\begin{equation}
    \rho(r)
    =
    \iiint F({\cal{E}},L)\,\txd\bfv
\label{detdf}
\end{equation}
for $F({\cal{E}},L)$. Hereby we have to take into account that not
every function $F({\cal{E}},L)$ that satisfies this equation is a
physically acceptable solution: an acceptable solution has to be
non-negative over the entire phase space. In general, the problem
of solving the integral equation (\ref{detdf}) is a degenerate
problem, because there are in general infinitely many distribution
functions possible for a given potential-density pair.

Particularly interesting are models for which the distribution
function and its moments can be computed analytically. Such models
have many useful applications, which can roughly be divided into
two classes. On the one hand, they can improve our general
understanding of physical processes in galaxies in an elegant way.
For example, they can serve as simple galaxy models, in which it
is easy to generate the starting conditions for $N$-body or Monte
Carlo simulations, or to test new data reduction or dynamical
modelling techniques. A quick look at the overwhelming success of
simple analytical models, such as the Plummer sphere (Plummer
1911; Dejonghe 1987), the isochrone sphere (H\'enon 1959, 1960),
the Jaffe model (Jaffe 1983) and the Hernquist model (Hernquist
1990), provides enough evidence. On the other hand, analytical
models are also useful for the detailed dynamical modelling of
galaxies. For example, in modelling techniques such as the QP
technique (Dejonghe 1989), a dynamical model for an observed
galaxy is built up as a linear combination of components, for each
of which the distribution function and its moments are known
analytically. As a result, the distribution function and the
moments of the final model are also analytical, which obviously
has a number of advantages.

Unfortunately, the number of dynamical models for which the
distribution function is known analytically is rather modest.
Moreover, most of them consist of distribution functions that are
isotropic or of the Osipkov-Merritt type, and therefore basically
one-dimensional. An exception is the completely analytical family
of anisotropic models described by Dejonghe (1987). These models
self-consistently generate the Plummer potential-density pair, a
simple yet useful model for systems with a constant density core.

During the last decade, however, it has become clear that, at
small radii, elliptical galaxies usually have central density
profiles that behave as $r^{-\gamma}$ with $0\leq\gamma\leq2.5$
(Lauer et al.\ 1995; Gebhardt et al. 1996). Such galaxies can
obviously not be adequately modelled with a constant density core.
This has stimulated the quest for simple potential-density pairs,
and corresponding distribution functions, with a central density
cusp. The first effort to construct such models was undertaken by
Ciotti (1991) and Ciotti \& Lanzoni (1997), who discussed the the
dynamical structure of stellar systems following the $R^{1/m}$ law
(S\'ersic 1968), a natural generalization of the empirical
$R^{1/4}$ law of de Vaucouleurs (1948). A major drawback of this
family, however, is that the spatial density and the distribution
function can not be written in terms of elementary functions (see
Mazure \& Capelato 2002). A more useful family is formed by the
so-called $\gamma$-models (Dehnen 1993; Tremaine et al.\ 1994),
characterized by a density proportional to $r^{-4}$ at large radii
and a divergence in the center as $r^{-\gamma}$ with
$0\leq\gamma\leq3$. The dynamical structure of models with this
potential-density pair has been extensively investigated (e.g.\
Carollo, de Zeeuw \& van der Marel 1995; Ciotti 1996; Meza \&
Zamorano 1997), but only for isotropic or Osipkov-Merritt type
distribution functions. Simple analytical models with a more
general anisotropy structure are still lacking.

In this paper we construct a number of families of completely
analytical anisotropic dynamical models that self-consistently
generate the Hernquist (1990) potential-density pair. It is a
special case of the family of $\gamma$-models, corresponding to
$\gamma=1$. In dimensionless units, the Hernquist
potential-density pair is given by
\begin{subequations}
\begin{gather}
    \psi(r)
    =
    \frac{1}{1+r}
\label{hernpot}
    \\
    \rho(r)
    =
    \frac{1}{2\pi}\,\frac{1}{r(1+r)^3}.
\label{hernrho}
\end{gather}
\end{subequations}
As the density diverges as $1/r$ for $r\rightarrow0$, the surface
brightness $I(R)$ will diverge logarithmically for
$R\rightarrow0$. More precisely, the surface brightness profile
has the form
\begin{equation}
    I(R)
    =
    \frac{1}{2\pi}\,
    \frac{(2+R^2)\,X(R)-3}{(1-R^2)^2},
\end{equation}
with $X(R)$ a continuous function defined as
\begin{equation}
    X(R)
    =
    \begin{cases}
    \,\,(1-R^2)^{-1/2}\,\arcsech R
    &\qquad\text{for $0\leq R\leq1$,} \\
    \,\,(R^2-1)^{-1/2}\,\arcsecans R
    &\qquad\text{for $1\leq R\leq\infty$.} \\
    \end{cases}
\label{defX}
\end{equation}

The paper is organized as follows. The general theory on the
inversion the fundamental integral equation (\ref{detdf}) is
resumed in Section 2. Each of the subsequent sections is devoted
to special cases of this inversion technique and the corresponding
family of Hernquist models. Isotropic models are the most simple
ones; Hernquist (1990) showed that, in the special case of
isotropy, the distribution function and its moments can be
calculated analytically. We repeat the most important
characteristics of the isotropic Hernquist model in Section 3. In
Section 4 we construct a one-parameter family of models with a
constant anisotropy. In Section 5, a two-parameter family of
Hernquist models is constructed by means of the Cuddeford (1991)
inversion technique. These models have an arbitrary anisotropy in
the center and are radially anisotropic at large radii. On the
contrary, in Section 6, a two-parameter family is constructed that
has a decreasing anisotropy profile, with arbitrary values for the
anisotropy in the center and the outer halo. Finally, Section 7
sums up.

\section{The construction of anisotropic models}

A general discussion on the inversion of the fundamental equation
(\ref{detdf}), and hence on the construction anisotropic
distribution functions for a given spherical potential-density
pair, is presented by Dejonghe (1986). The key ingredient of the
inversion procedure is the concept of the augmented mass density
$\tilde{\rho}(\psi,r)$, which is a function of potential and
radius, such that the condition
\begin{equation}
    \tilde{\rho}(\psi(r),r)
    \equiv
    \rho(r)
\label{condrho}
\end{equation}
is satisfied. The augmented mass density is in fact equivalent to
the distribution function $F({\cal{E}},L)$: with every augmented
density $\tilde{\rho}(\psi,r)$ we can associate a distribution
function $F({\cal{E}},L)$ and vice versa. There exist various
transition formulae between these two equivalent forms of a
dynamical model, amongst others a formalism that uses combined
Laplace-Mellin transforms.

Besides providing a nice way to generate a distribution function
for a given potential-density pair, the augmented density is also
very useful to calculate the moments of the distribution function.
The anisotropic moments are defined as
\begin{equation}
    \mu_{2n,2m}(r)
    =
    2\pi\iint F({\cal{E}},L)\,v_r^{2n}\,v_t^{2m+1}\,\txd v_r\,\txd v_t,
\label{defgenanimom}
\end{equation}
where $v_t\equiv\sqrt{v_\theta^2+v_\phi^2}$ is the transverse
velocity. One can derive a relation that links the higher-order
moments to the augmented mass density $\tilde{\rho}$, when written
explicitly as a function of $\psi$ and $r$,
\begin{multline}
    \tilde{\mu}_{2n,2m}(\psi,r)
    =
    \frac{2^{m+n}}{\sqrt{\pi}}\,
    \frac{\Gamma(n+\tfrac{1}{2})}{\Gamma(m+n)}\,
    \\
    \times
    \int_0^\psi
    (\psi-\psi')^{m+n-1}\,
    \frac{\txd^{m}}{(\txd r^2)^{m}}
    \left[
    r^{2m}\,\tilde{\rho}(\psi',r)
    \right]
    \txd\psi'.
\label{genmom}
\end{multline}
In particular, the radial and transverse velocity dispersions can
be found from the density through the relations,
\begin{subequations}
\begin{gather}
    \sigma_r^2(r)
    =
    \frac{1}{\rho(r)}
    \int_0^{\psi(r)} \tilde{\rho}(\psi',r)\,\txd\psi',
\label{gensigr:def}
    \\
    \sigma_t^2(r)
    =
    \frac{2}{\rho(r)}
    \int_0^{\psi(r)}
    \frac{\txd}{\txd r^2}
    \left[r^2\,\tilde{\rho}(\psi',r)\right]
    \txd\psi'.
\label{gensigt:def}
\end{gather}
\end{subequations}
By means of these functions, we can define the anisotropy
$\beta(r)$ as
\begin{equation}
    \beta(r)
    =
    1-\frac{\sigma_t^2(r)}{2\sigma_r^2(r)}.
\label{defbeta}
\end{equation}
We will in this paper consider augmented densities which are
separable functions of $\psi$ and $r$, and we introduce the
notation
\begin{equation}
    \tilde{\rho}(\psi,r)
    =
    f(\psi)\,g(r).
\label{splitaugdens}
\end{equation}
For such models, the anisotropy can be directly calculated from
the augmented density as
\begin{equation}
    \beta(r)
    =
    1-\frac{1}{g(r)}\,\frac{\txd}{\txd r^2}
    \left[r^2\,g(r)\right],
\label{betasplits}
\end{equation}
as a result of the formulae (\ref{gensigr:def}b), (\ref{defbeta})
and (\ref{splitaugdens}).

\section{Isotropic models}

\subsection{Background}

The simplest dynamical models are those where the augmented
density is a function of the potential only,
$\tilde{\rho}=\tilde{\rho}(\psi)$. For such models, the
distribution function is only a function of the binding energy,
i.e.\ the distribution function is isotropic. In this case, the
integral equation (\ref{detdf}) can be inverted to find the
well-known Eddington relation
\begin{equation}
    F({\cal{E}})
    =
    \frac{1}{2\sqrt{2}\pi^2}
    \frac{\txd}{\txd {\cal{E}}}
    \int_0^{\cal{E}}
    \frac{\txd\tilde{\rho}}{\txd\psi}\,
    \frac{\txd\psi}{\sqrt{{\cal{E}}-\psi}}.
\label{eddington}
\end{equation}
For such isotropic models, we do not use the general anisotropic
moments (\ref{defgenanimom}), but define the isotropic moments as
\begin{equation}
    \mu_{2n}(r)
    =
    4\pi\int F({\cal{E}})\,v^{2n+2}\,\txd v.
\end{equation}
Similarly as for the anisotropic moments, we can derive a relation
that allows to calculate the augmented isotropic moments from the
augmented density $\tilde{\rho}(\psi)$. Indeed, they satisfy the
relation (Dejonghe 1986)
\begin{equation}
    \tilde{\mu}_{2n}(\psi)
    =
    \frac{(2n+1)!!}{(n-1)!!}
    \int_0^{\psi}
    (\psi-\psi')^{n-1}\,\tilde{\rho}(\psi')\,\txd\psi'.
\label{calcisomom}
\end{equation}
In particular, we obtain an expression for the velocity dispersion
profile by setting $n=1$,
\begin{equation}
    \sigma^2(r)
    =
    \frac{1}{\rho(r)}
    \int_0^{\psi(r)}\tilde{\rho}(\psi')\,\txd\psi'.
\label{isodig:def}
\end{equation}

\subsection{The isotropic Hernquist model}

The isotropic model that corresponds to the potential-density pair
(\ref{hernpot}b) is described in full detail by Hernquist (1990).
We restrict ourselves by resuming the most important results, for
a comparison with the anisotropic models discussed later in this
paper. The augmented density reads
\begin{equation}
    \tilde{\rho}(\psi)
    =
    \frac{1}{2\pi}\,\frac{\psi^4}{1-\psi}.
\label{ell_iso}
\end{equation}
Substituting this density into Eddington's formula
(\ref{eddington}) yields the distribution function
\begin{multline}
    F({\cal{E}})
    =
    \frac{1}{8\sqrt{2}\pi^3}
    \\ \times
    \left[
    \frac{\sqrt{{\cal{E}}}\,(1-2{\cal{E}})\,(8{\cal{E}}^2-8{\cal{E}}-3)}{(1-{\cal{E}})^2}
    +
    \frac{3\arcsin\sqrt{{\cal{E}}}}{(1-{\cal{E}})^{5/2}}
    \right].
\label{df_iso}
\end{multline}
Combining the density (\ref{ell_iso}) with the general formula
(\ref{calcisomom}) gives us the moments of the distribution
function,
\begin{equation}
    \tilde{\mu}_{2n}(\psi)
    =
    \frac{3\cdot2^{n+9/2}}{(2\pi)^{3/2}}\,
    \frac{\Gamma(n+\tfrac{1}{2})}{\Gamma(n+5)}\,
    \psi^{n+4}\,
    {}_2F_1\left(5,1;n+5;\psi\right).
\end{equation}
For all $n\geq0$, this expression can be written in terms of
rational functions and logarithms. For example, for the velocity
dispersions, we obtain after substitution of the Hernquist
potential (\ref{hernpot}),
\begin{equation}
    \sigma^2(r)
    =
    r\,(1+r)^3\,\ln\left(\frac{1+r}{r}\right)
    -\frac{r\,(25+52r+42r^2+12r^3)}{12(1+r)},
\label{disp_iso}
\end{equation}
in agreement with equation (10) of Hernquist (1990). From an
observational point of view, it is very useful to obtain an
explicit expression for the line-of-sight velocity dispersion. For
isotropic models, the line-of-sight dispersion is easily found by
projecting the second-order moment on the plane of the sky, i.e.\
\begin{equation}
    \sigma_p^2(R)
    =
    \frac{2}{I(R)}\,
    \int_R^\infty
    \frac{\rho(r)\,\sigma^2(r)\,r\,\txd r}{\sqrt{r^2-R^2}}.
\label{calcsigpiso}
\end{equation}
For the Hernquist model this yields after some algebra
\begin{multline}
    I(R)\,\sigma_p^2(R)
    =
    \frac{1}{24\pi\,(1-R^2)^3}
    \\
    \times
    \Bigl[3R^2\,(20-35R^2+28R^4-8R^6)\,X(R)
    \\
    +
    (6-65R^2+68R^4-24R^6)\Bigr]
    -\frac{R}{2}.
\end{multline}

\section{Models with a constant anisotropy}

\subsection{Background}
\label{canitheory.sec}

A special family of distribution functions that can easily be
generated using the technique outlined in Section 2 corresponds to
models with a density that depends on $r$ only through a factor
$r^{-2\beta}$, i.e.\
\begin{equation}
    \tilde{\rho}(\psi,r)
    =
    f(\psi)\,r^{-2\beta}.
\end{equation}
It is well known that such densities correspond to models with a
constant anisotropy (i.e.\ Binney \& Tremaine 1987), which can
easily be checked by introducing $g(r)=r^{-2\beta}$ in the formula
(\ref{betasplits}). For a given potential-density pair, a family
of models with constant anisotropy can hence be constructed by
inverting the potential as $r(\psi)$, and defining
\begin{equation}
    f(\psi)
    =
    \rho(\psi)\,\bigl(r(\psi)\bigr)^{2\beta}.
\label{constructcani}
\end{equation}
Notice that such models are not the only constant anisotropy
models corresponding to a given potential-density pair, as argued
section 1.5 of Dejonghe (1986). This family, however, is very
attractive due to its relative simpleness. In particular, the
corresponding distribution function is a power law in $L$, and can
be found through an Eddington-like formula,
\begin{multline}
    F({\cal{E}},L)
    =
    \frac{2^\beta}{(2\pi)^{3/2}}\,
    \frac{L^{-2\beta}}{\Gamma(1-\beta)\,\Gamma(\tfrac{1}{2}+\beta)}\,
    \\
    \times
    \frac{\txd}{\txd {\cal{E}}}
    \int_0^{\cal{E}}
    \frac{\txd f}{\txd\psi}\,\frac{\txd\psi}{({\cal{E}}-\psi)^{1/2-\beta}}.
\label{dfcani}
\end{multline}
For the moments of the distribution function, the relation
(\ref{genmom}) can be simplified to
\begin{multline}
    \tilde{\mu}_{2n,2m}(\psi,r)
    =
    \frac{2^{m+n}}{\sqrt{\pi}}\,
    \frac{\Gamma(n+\tfrac{1}{2})\,\Gamma(m+1-\beta)}{\Gamma(m+n)\,\Gamma(1-\beta)}\,
    r^{-2\beta}
    \\
    \times
    \int_0^\psi
    (\psi-\psi')^{m+n-1}\,
    f(\psi')\,
    \txd\psi'.
\end{multline}
In particular, the radial velocity dispersions reads
\begin{equation}
    \sigma_r^2(r)
    =
    \frac{1}{f(\psi(r))}
    \int_0^{\psi(r)} f(\psi')\,\txd\psi'.
\label{canisig:def}
\end{equation}

\subsection{Hernquist models with a constant anisotropy}
\label{cani.sec}

\subsubsection{The distribution function}
\label{canidf.sec}

\begin{figure*}
\centering
\includegraphics[clip]{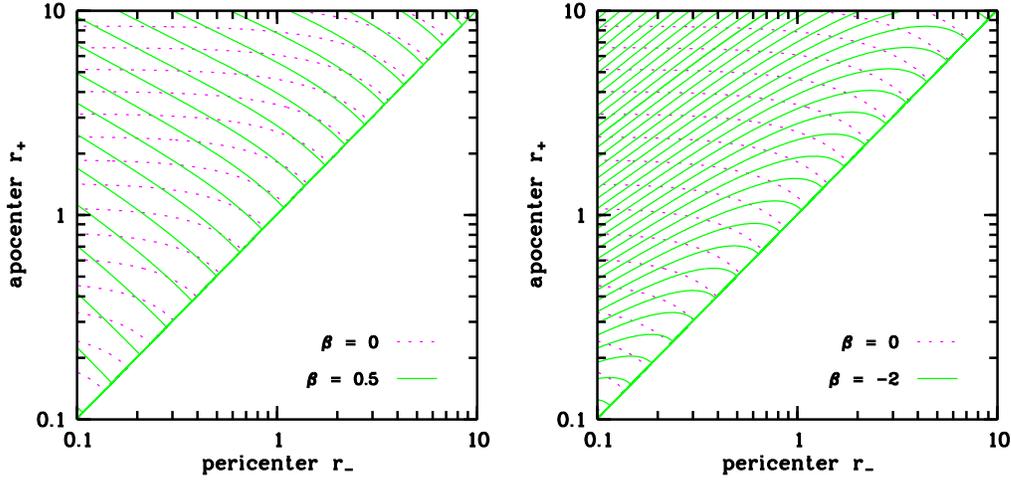}
\caption{The distribution function of the Hernquist models with a
constant anisotropy, represented as isoprobability contours in
turning point space. The distribution functions in solid lines
represent a radial model with $\beta=\tfrac{1}{2}$ (left panel)
and a tangential model with $\beta=-2$ (right panel). The dotted
contour lines in both panels correspond to the isotropic Hernquist
model.} \label{MS2799f1.eps}
\end{figure*}

Applying the formula (\ref{constructcani}) to the Hernquist
potential-density pair (\ref{hernpot}b) yields
\begin{equation}
    f(\psi)
    =
    \frac{1}{2\pi}\,
    \psi^{4-2\beta}\,(1-\psi)^{2\beta-1}.
\label{ell_cani}
\end{equation}
Substituting this expression into the general formula
(\ref{dfcani}) gives us the corresponding distribution function
\begin{multline}
    F({\cal{E}},L)
    =
    \frac{2^\beta}{(2\pi)^{5/2}}\,
    \frac{\Gamma(5-2\beta)}{\Gamma(1-\beta)\,\Gamma(\tfrac{7}{2}-\beta)}\,
    \\
    \times
    L^{-2\beta}\,{\cal{E}}^{\tfrac{5}{2}-\beta}\,
    {}_2F_1\left(5-2\beta,1-2\beta;\frac{7}{2}-\beta;{\cal{E}}\right).
\label{df_cani}
\end{multline}
This expression reduces to the isotropic distribution function
(\ref{df_iso}) for $\beta=0$, as required. Whether the expression
(\ref{df_cani}) corresponds to a physically acceptable
distribution function for a given value of $\beta$ depends on the
condition that the distribution function has to be positive over
the entire phase space.

It is no surprise that the distribution function is not positive
for the highest possible values of $\beta$, because models where
only the radial orbits are populated can only be supported by a
density profile that diverges as $r^{-2}$ or steeper in the center
(Richstone \& Tremaine 1984). It turns out that the distribution
function (\ref{df_cani}) is everywhere non-negative for
$\beta\leq\tfrac{1}{2}$.

For all integer and half-integer values of $\beta$, the
hypergeometric series in (\ref{df_cani}) can be expressed in terms
of elementary functions. Very useful are half-integer values of
$\beta$, because the energy-dependent part of the distribution
function can then be written as a rational function of
${\cal{E}}$. For integer values of $\beta$, the hypergeometric
series can be written as a function containing integer and
half-integer powers of ${\cal{E}}$ and $1-{\cal{E}}$ and a factor
$\arcsin\sqrt{{\cal{E}}}$, similar to the isotropic distribution
function (\ref{df_iso}).

The limiting model $\beta=\tfrac{1}{2}$ is particularly simple. It
has an augmented density that is a power law of potential and
radius,
\begin{equation}
    \tilde{\rho}(\psi,r)
    =
    \frac{1}{2\pi}\,\frac{\psi^3}{r},
\end{equation}
and the corresponding distribution function simply reads on 1973),
\begin{equation}
    F({\cal{E}},L)
    =
    \frac{3}{4\pi^3}\,\frac{{\cal{E}}^2}{L}.
\label{Fcanirad}
\end{equation}
This model is a special case of generalized polytropes discussed
by Fricke (1951) and H\'enon (1973).

In Fig.\ \ref{MS2799f1.eps} we compare the distribution functions
of the radial model with $\beta=\tfrac{1}{2}$ and the tangential
model with $\beta=-2$ with the distribution function of the
isotropic Hernquist model. The distribution functions are shown by
means of their isoprobability contours in turning point space,
which can easily be interpreted in terms of orbits. Compared to
the isotropic model, the radial model prefers orbits on the upper
left side of the diagram, with an apocenter much larger than the
pericenter, i.e.\ elongated orbits. The isoprobability contours of
tangential models on the other hand lean towards the diagonal axes
where pericenter and apocenter are equal, i.e.\ nearly-circular
orbits are preferred.

\subsubsection{The velocity dispersions}

\begin{figure}
\centering
\includegraphics[clip]{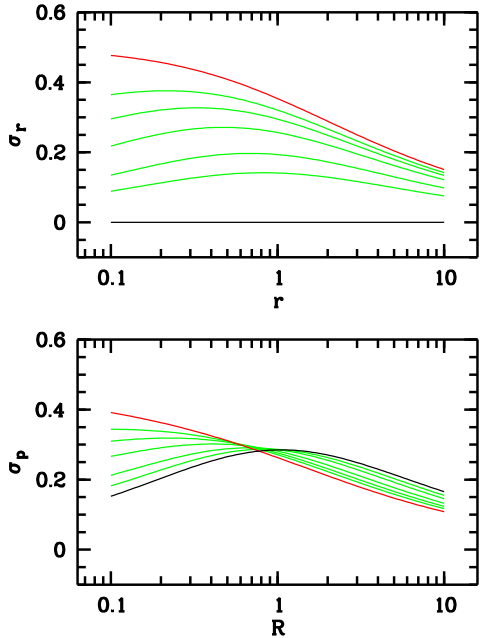}
\caption{The velocity dispersion of the Hernquist models with a
constant anisotropy. The upper and lower panels show the radial
velocity dispersions $\sigma_r(r)$ and the line-of-sight velocity
dispersion $\sigma_p(R)$. The profiles are shown for different
values of the anisotropy parameter $\beta$: plotted are
$\beta=\tfrac{1}{2}$, $\tfrac{1}{4}$, 0, $-\tfrac{1}{2}$, $-2$,
$-5$ and the limit case $\beta\rightarrow-\infty$.}
\label{MS2799f2.eps}
\end{figure}

By means of substituting the expression (\ref{ell_cani}) into the
general formula (\ref{genmom}), we can derive an analytical
expression for all moments of the distribution function,
\begin{multline}
    \tilde{\mu}_{2n,2m}(\psi,r)
    =
    \frac{2^{n+m-1}}{\pi^{3/2}}\,
    \frac{\Gamma(5-2\beta)\,\Gamma(n+\tfrac{1}{2})\,\Gamma(m+1-\beta)}
    {\Gamma(m+n+5-2\beta)\,\Gamma(1-\beta)}
    \\ \qquad\quad\times
    r^{-2\beta}\,\psi^{m+n+4-\beta}\,
    \\ \times
    {}_2F_1\left(5-2\beta,1-2\beta;m+n+5-2\beta;\psi\right).
\end{multline}
We are mainly interested in the velocity dispersions, which can be
conveniently written by means of the incomplete Beta function
(Abramowitz \& Stegun 1972),
\begin{equation}
    \sigma_r^2(r)
    =
    r^{1-2\beta}\,(1+r)^3\,
    \text{B}_{\frac{1}{1+r}}\Bigl(5-2\beta,2\beta\Bigr).
\label{canisig}
\end{equation}
For all anisotropies $\beta<\tfrac{1}{2}$, the radial dispersion
equals zero in the center of the galaxy, rises until a maximum and
then decreases again towards zero for $r\rightarrow\infty$. The
asymptotic behavior for $r\gg1$ is
\begin{equation}
    \sigma_r^2(r)
    \approx
    \frac{1}{5-2\beta}\,\frac{1}{r}
    +\cdots
\label{sigrexp}
\end{equation}
The expression (\ref{canisig}) can be written in terms of
elementary functions for all $\beta$ with $4\beta$ an integer. For
the integer and half-integer values of $\beta$, the expression
involves polynomials in $r$ and a factor $\ln(1+1/r)$, very
analogous to the expression (\ref{disp_iso}) of the isotropic
Hernquist model. For the quarter-integer values of $\beta$, it
contains polynomials and square roots in $r$ and a factor
$\arccotg\sqrt{r}$.

Particular cases are the models that correspond to the most radial
and tangential distribution functions. On the one hand, the limit
case $\beta=\tfrac{1}{2}$ has the simple velocity dispersion
profiles
\begin{equation}
    \sigma_r^2(r)
    =
    \sigma_t^2(r)
    =
    \frac{1}{4}\,\frac{1}{1+r}.
\label{disp_cani}
\end{equation}
Particular about this dispersion profile is that it assumes a
finite value in the center. On the other side of the range for
possible anisotropies, we can consider the limit
$\beta\rightarrow-\infty$, which corresponds to a model with
purely circular orbits. For such a model, the radial dispersion is
of course identically zero, whereas the transverse velocity
dispersion is the circular velocity corresponding to the Hernquist
potential,
\begin{equation}
    \sigma_t^2(r)
    =
    v_c^2(r)
    =
    \frac{1}{2}\,\frac{r}{(1+r)^2}.
\end{equation}

In the top panel of Fig.\ \ref{MS2799f2.eps} we plot the radial
velocity dispersion profile for various models with a different
anisotropy $\beta$. Both at small and large radii, the radial
dispersion is a decreasing function of $\beta$, as expected.

The line-of-sight velocity dispersion for anisotropic models is
found through the formula
\begin{equation}
    \sigma_p^2(R)
    =
    \frac{2}{I(R)}\,
    \int_R^\infty
    \frac{\rho(r)\,\sigma_{\text{los}}^2(r,R)\,r\,\txd r}{\sqrt{r^2-R^2}},
\end{equation}
where $\sigma_{\text{los}}(r,R)$ is the velocity dispersion at the
position $r$ on the line of sight $R$ in the direction of
observer. It is a linear combination of the radial and transverse
velocity dispersions in this point,
\begin{equation}
    \sigma_{\text{los}}^2(r,R)
    =
    \left(1-\frac{R^2}{r^2}\right)\sigma_r^2(r)
    + \frac{R^2}{2r^2}\,\sigma_t^2(r).
\label{sigmalos}
\end{equation}
We can equivalently write
\begin{equation}
    \sigma_p^2(R)
    =
    \frac{2}{I(R)}\,
    \int_R^\infty
    \left[1-\beta(r)\frac{R^2}{r^2}\right]\,
    \frac{\rho(r)\,\sigma_r^2(r)\,r\,\txd r}{\sqrt{r^2-R^2}}.
\label{calcsigp}
\end{equation}
For the general Hernquist models with a constant anisotropy, the
integration (\ref{calcsigp}) cannot be performed analytically. But
for all integer and half-integer $\beta$'s, $\sigma_p^2(R)$ can be
expressed in terms of polynomials and the function $X$, defined in
equation (\ref{defX}). For example, for the limit model
$\beta=\tfrac{1}{2}$ we obtain after some algebra
\begin{multline}
    I(R)\,\sigma_p^2(R)
    =
    \frac{1}{48\pi\,(1-R^2)^3}
    \\ \times
    \Bigl[3(4-14R^2+35R^4-28R^6+8R^8)\,X(R)
    \\ -
    (28-57R^2+68R^4-24R^6)\Bigr]
    +\frac{R}{4}.
\end{multline}
For the other limit model, the one with only circular orbits, we
find
\begin{multline}
    I(R)\,\sigma_p^2(R)
    =
    \frac{R^2}{48\pi\,(1-R^2)^4}
    \\ \times
    \Bigl[-(120-120R^2+189R^4-108R^6+24R^8)\,X(R)
    \\ +
    (154-117R^2+92R^4-24R^6)\Bigr]
    +\frac{R}{4},
\end{multline}
in agreement with Hernquist (1990).

In the bottom panel of Fig.\ \ref{MS2799f2.eps} we plot the
line-of-sight dispersion profiles for a number of different values
of $\beta$. The behavior of the individual profiles is analogous
to the spatial dispersion profiles: except for the
$\beta=\tfrac{1}{2}$ model, which has a finite central dispersion,
the $\sigma_p$ profiles start at zero in the center, rise strongly
until a certain maximum and then decrease smoothly towards zero at
large projected radii. The behavior for $R\gg1$ can be quantified
if we introduce the asymptotic expansion (\ref{sigrexp}) into the
formula (\ref{calcsigp}),
\begin{equation}
    \sigma_p^2(R)
    \approx
    \frac{8}{15\pi}\left(\frac{5-4\beta}{5-2\beta}\right)\frac{1}{R}
    +
    \cdots.
\label{sigpexp}
\end{equation}
The dependence of the line-of-sight dispersion as a function of
the anisotropy is depends strongly on the projected radius: at
small projected radii, $\sigma_p$ decreases with increasing
$\beta$, whereas for large radii, $\sigma_p$ increases with
increasing $\beta$. In other words, radial models have a larger
central and a smaller outer line-of-sight dispersion than their
tangential counterparts. This is a direct consequence of the
weight of the radial and transversal velocity dispersions in the
linear combination (\ref{sigmalos}): at small projected radii, the
radial dispersion contributes the dominant term, whereas for the
outer lines of sight, the transversal dispersion term dominates.

\section{Models with increasing anisotropy}
\label{cudd.sec}

\subsection{Background}

Osipkov (1979) and Merritt (1985) developed an inversion technique
for a special class of distribution functions that only depend on
energy and angular momentum through the combination
\begin{equation}
    Q\equiv {\cal{E}}-\frac{L^2}{2r_a^2},
\end{equation}
with $r_a$ the so-called anisotropy radius, with the additional
condition that
\begin{equation}
    F({\cal{E}},L)=0
    \qquad\quad
    \text{for $Q<0$}.
\label{Qcond}
\end{equation}
Such models correspond to an augmented density of the form
\begin{equation}
    \tilde{\rho}(\psi,r)
    =
    \left(1+\frac{r^2}{r_a^2}\right)^{-1}
    f(\psi).
\end{equation}
In this case, the fundamental integral equation (\ref{detdf}) can
be inverted in a similar way as the Eddington relation,
\begin{equation}
    F({\cal{E}},L)
    =
    \frac{1}{2\sqrt{2}\,\pi^2}\,
    \frac{\txd}{\txd Q}
    \int_0^Q
    \frac{\txd f}{\txd\psi}\,
    \frac{\txd\psi}{\sqrt{Q-\psi}}.
\end{equation}
The anisotropy $\beta(r)$ for the Osipkov-Merritt models can be
found by means of formula (\ref{betasplits}),
\begin{equation}
    \beta(r)
    =
    \frac{r^2}{r^2+r_a^2}.
\end{equation}
These models are hence isotropic in the center and completely
radially anisotropic in the outer regions. The parameter $r_a$
determines how soon the anisotropy turns from isotropic to radial.
In particular, for $r_a\rightarrow\infty$, $Q$ is nothing else
than the binding energy, and the Osipkov-Merritt models reduce to
the isotropic models.

The Osipkov-Merritt models were generalized by Cuddeford (1991),
who considered models which correspond to an augmented density of
the form
\begin{equation}
    \tilde{\rho}(\psi,r)
    =
    r^{-2\beta_0}
    \left(1+\frac{r^2}{r_a^2}\right)^{-1+\beta_0}
    f(\psi).
\end{equation}
These models reduce to the Osipkov-Merritt models if we set
$\beta_0=0$. Within this formalism, the distribution function can
be calculated in a similar way as for the Osipkov-Merritt models.
The distribution functions have the general form
\begin{equation}
    F({\cal{E}},L)
    =
    F_0(Q)\,L^{-2\beta_0},
\end{equation}
with the additional condition (\ref{Qcond}). The solution of the
integral equation (\ref{detdf}), valid for $\beta_0<1$, reads in
this case
\begin{subequations}
\begin{multline}
    F({\cal{E}},L)
    =
    \frac{2^{\beta_0}}{(2\pi)^{3/2}}\,
    \frac{L^{-2\beta_0}}{\Gamma(1-\beta_0)\,\Gamma(1-\theta)}
    \\ \times
    \frac{\txd}{\txd Q}
    \int_0^Q
    \frac{\txd^m f}{\txd\psi^m}\,
    \frac{\txd\psi}{(Q-\psi)^\theta},
\label{F0gen}
\end{multline}
where
\begin{gather}
    m = 1+\intoperator\left(\frac{1}{2}-\beta_0\right)
    \\
    \theta = \fracoperator\left(\frac{1}{2}-\beta_0\right).
\end{gather}
\end{subequations}
The most interesting cases are those where $\beta_0$ is either
integer or half-integer. For integer values of $\beta_0$, the
general formula (\ref{F0gen}bc) reduces to
\begin{equation}
    F({\cal{E}},L)
    =
    \frac{2^{\beta_0}}{2\sqrt{2}\,\pi^2}\,
    \frac{L^{-2\beta_0}}{\Gamma(1-\beta_0)}\,
    \frac{\txd}{\txd Q}
    \int_0^Q
    \frac{\txd^{1-\beta_0}f}{\txd\psi^{1-\beta_0}}\,
    \frac{\txd\psi}{\sqrt{Q-\psi}}.
\label{F0int}
\end{equation}
For half-integer values of $\beta_0$, the integral equation
(\ref{detdf}) is a degenerate integral equation, which can be
solved without a single integration (Dejonghe 1986; Cuddeford
1991),\footnote{Apparently, a factor
$\Gamma(\tfrac{3}{2}-\beta_0)$ is missing in the denominator in
formula (30) of Cuddeford (1991).}
\begin{equation}
    F({\cal{E}},L)
    =
    \frac{2^{\beta_0}}{(2\pi)^{3/2}}\,
    \frac{L^{-2\beta_0}}{\Gamma(1-\beta_0)}\,
    \left[
    \frac{\txd^{\frac{3}{2}-\beta_0}f}{\txd\psi^{\frac{3}{2}-\beta_0}}
    \right]_{\psi=Q}.
\label{F0odd}
\end{equation}
As for the Osipkov-Merritt models, the anisotropy of the Cuddeford
models has a very simple functional form, which can be found
through (\ref{betasplits}),
\begin{equation}
    \beta(r)
    =
    \frac{r^2+\beta_0r_a^2}{r^2+r_a^2}.
\label{cuddanis}
\end{equation}
They hence have an anisotropy $\beta_0$ in the
center,\footnote{This explains why we prefer the parameter
$\beta_0$ above $\alpha=-\beta_0$ originally adopted by Cuddeford
(1991).} and become completely radially anisotropic in the outer
regions. The anisotropy radius $r_a$ is again a degree for how
quick this transition takes place. In particular, for
$r_a\rightarrow\infty$, the Cuddeford models reduce to models with
a constant anisotropy $\beta=\beta_0$. Because the range of values
for $\beta_0$ for which the inversion (\ref{F0gen}bc) is
mathematically defined is only restricted by $\beta_0<1$,
distribution functions can in principle be calculated with any
degree of anisotropy in the center, ranging from very radial to
extremely tangential. Whether these distribution functions
correspond to physically acceptable solutions depends on the
positivity, however.

\subsection{Hernquist models with increasing anisotropy}

\subsubsection{The distribution function}

For the Hernquist potential-density pair (\ref{hernpot}b), the
augmented density corresponding to the Cuddeford formalism is
readily calculated. We obtain
\begin{equation}
    f(\psi)
    =
    \frac{1}{2\pi}
    \left[
    1+\lambda\left(\frac{1-\psi}{\psi}\right)^2\right]^{1-\beta_0}
    \frac{\psi^{4-2\beta_0}}{(1-\psi)^{1-2\beta_0}},
\label{rho2hern2}
\end{equation}
where we have set $\lambda=1/r_a^2$. Combining this expression
with the general Cuddeford solution (\ref{F0gen}bc) we can obtain
distribution functions that self-consistently generate the
Hernquist potential-density pair, and which have an arbitrary
anisotropy in the center and a completely radial structure in the
outer regions. In order to represent physically acceptable
dynamical models, it is necessary that these distribution
functions are positive over the entire phase space, i.e.\
$F({\cal{E}},L)\geq0$ for $0\leq Q\leq1$. Before trying to
actually calculate the distribution functions, it is useful to
investigate which region in the $(\beta_0,\lambda)$ parameter
space corresponds to physically acceptable distribution functions.

First of all, it is obvious that the models with
$\beta_0>\tfrac{1}{2}$ will not correspond to non-negative
distribution functions: the distribution function is already too
radial for $\lambda=0$ (Sec.\ \ref{canidf.sec}), and will become
even more radial for larger $\lambda$. We can therefore limit the
subsequent discussion to $\beta_0\leq\tfrac{1}{2}$. Now consider
such a fixed value $\beta_0$, and consider all Cuddeford models
corresponding to this central anisotropy. For
$\lambda\rightarrow0$, the Cuddeford model reduces to the model
with constant anisotropy $\beta_0$, which is physically acceptable
(Sec.\ \ref{canidf.sec}). For $\lambda\rightarrow\infty$, the
distribution function will only consist of radial orbits, for
which the distribution function is not positive. It can therefore
be expected that, for a given value of $\beta_0\leq\tfrac{1}{2}$,
a range of $\lambda$'s is allowed, starting from 0 up to a certain
$\lambda_{\text{max}}$.

\begin{table}
\centering \caption{The range of anisotropy radii which give rise
to a positive distribution function of the Cuddeford type,
consistent with the Hernquist potential-density pair. For a given
value of $\beta_0$, this range corresponds to
$0\leq\lambda\leq\lambda_{\text{max}}$, or equivalently, to
$r_{a,\text{min}}\leq r_a\leq\infty$.} \label{lambdamax.tab}
\begin{tabular}{rcc} \hline
$\beta_0$\quad{} & $\lambda_{\text{max}}$ & $r_{a,\text{min}}$ \\
\hline
$\leq-1.500$ & 0.000 & $\infty$ \\
$-1.375$ & 1.764 & 0.753 \\
$-1.250$ & 3.598 & 0.527 \\
$-1.125$ & 5.550 & 0.424 \\
$-1.000$ & 7.582 & 0.363 \\
$-0.875$ & 9.680 & 0.321 \\
$-0.750$ & 11.83 & 0.291 \\
$-0.625$ & 14.02 & 0.267 \\
$-0.500$ & 16.23 & 0.248 \\
$-0.375$ & 18.51 & 0.232 \\
$-0.250$ & 20.57 & 0.220 \\
$-0.125$ & 22.61 & 0.210 \\
$0.000$  & 24.42 & 0.202 \\
$0.125$  & 25.87 & 0.197 \\
$0.250$  & 26.70 & 0.194 \\
$0.375$  & 26.42 & 0.195 \\
$0.500$  & 24.00 & 0.204 \\ \hline
\end{tabular}
\end{table}

Next, we have to investigate how $\lambda_{\text{max}}$ varies
with $\beta_0$, i.e.\ which anisotropy radii are allowed for a
given central anisotropy ? Distribution functions with a strong
central tangential anisotropy and a small anisotropy radius are
likely to be negative. Indeed, consider the orbital structure of
such a galaxy. Because the outer regions of the galaxy $(r\gg
r_a)$ are strongly radially anisotropic, the vast majority of the
stars there must be on nearly radial orbits. These stars also pass
through the central regions, where they will contribute to the
central density and radial velocity dispersion as well. The
smaller the value of $r_a$, i.e.\ the larger the value of
$\lambda$, the stronger the contribution of stars on such nearly
radial orbits. In order to create a core where the anisotropy is
tangential, a large number of stars hence have to be added which
move on tightly bound nearly circular orbits. But we are limited
from keeping on adding such stars, because we cannot exceed the
spatial density of the Hernquist profile, which has only a fairly
weak $r^{-1}$ divergence. We therefore expect that no Cuddeford
models will exist beyond a certain minimal $\beta_0$ (except for
the degenerate case of the constant anisotropy models, which have
no radial anisotropy at large radii). Moreover, it can be expected
that for models with a tangential central anisotropy, the range of
anisotropy radii is more restricted than for models with a radial
or isotropic central anisotropy, i.e.\ that
$\lambda_{\text{max}}(\beta_0)$ is a increasing function of
$\beta_0$.

\begin{figure}
\centering
\includegraphics[clip,width=0.4\textwidth]{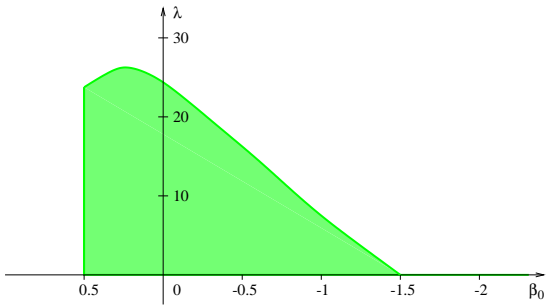}
\caption{The region in $(\beta_0,\lambda)$ space corresponding to
a positive distribution function of the Cuddeford type, consistent
with the Hernquist potential-density pair. } \label{MS2799f3.eps}
\end{figure}

\begin{figure*}
\centering
\includegraphics[clip]{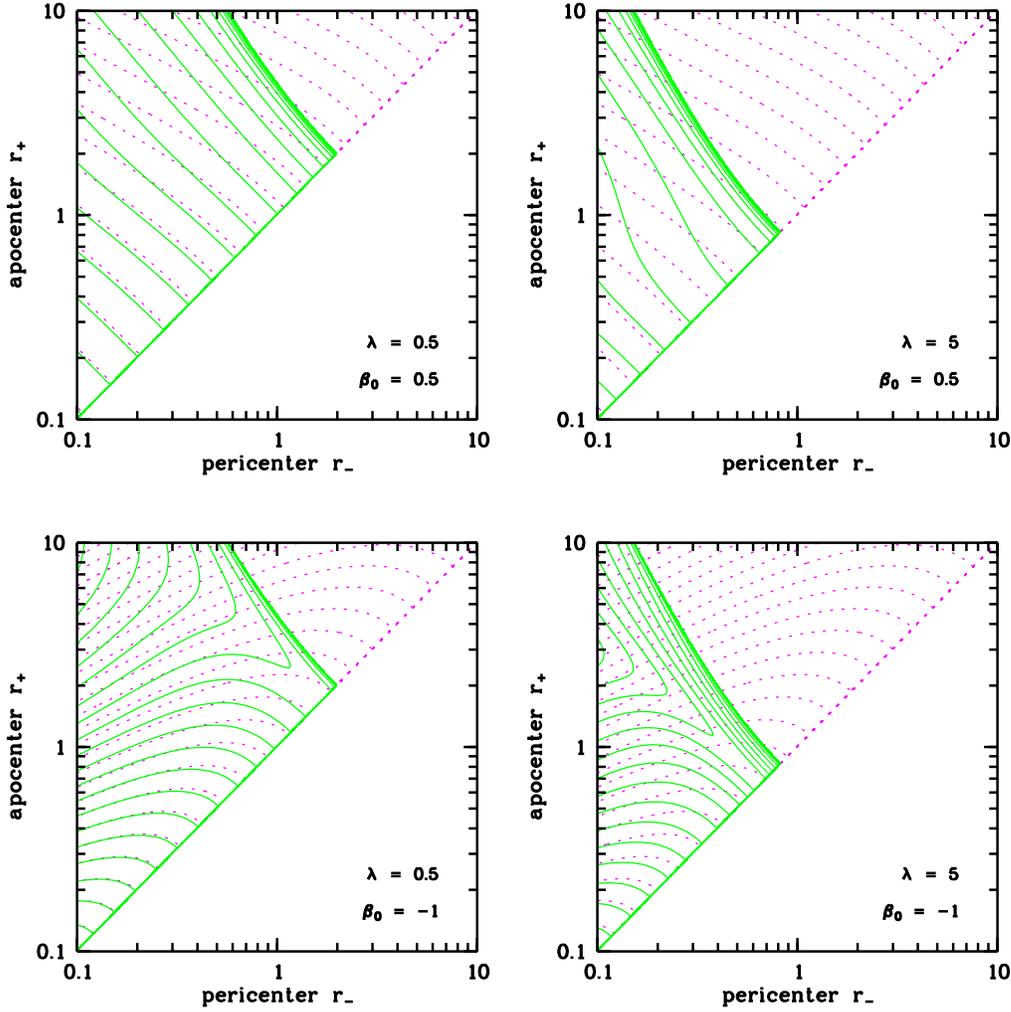}
\caption{Comparison of the distribution function corresponding to
Hernquist models of the Cuddeford type and Hernquist models with a
constant anisotropy. The distribution functions are represented as
isoprobability contours in turning point space. The solid lines
correspond to the Cuddeford distribution functions, with the
parameters $\beta_0$ and $\lambda$ displayed in the down left
corner of each diagram. The dotted lines represent the
distribution functions of the corresponding Hernquist models with
a constant anisotropy $\beta_0$.} \label{MS2799f4.eps}
\end{figure*}

By numerical evaluation of the integral in equation (\ref{F0gen}),
we calculated $\lambda_{\text{max}}(\beta_0)$ for a number of
values for $\beta_0$ (Table \ref{lambdamax.tab}). The region in
parameter space where the Cuddeford-Hernquist models are physical
is shown in Fig.\ \ref{MS2799f3.eps}. Notice that all models with
$\beta_0\leq-\tfrac{3}{2}$ and $\lambda>0$ are negative at some
point in phase space and are thus unphysical: the Hernquist
potential-density pair can support no (non-degenerate)
distribution functions of the Cuddeford type with a central
anisotropy $\beta_0\leq-\frac{3}{2}$.

We are primarily interested in those models where the distribution
function can be expressed in terms of elementary functions. This
is of course possible for all half-integer values of $\beta_0$,
because the calculation of the distribution function involves no
integrations. Also for the integer values of $\beta_0$, the
distribution function can be calculated analytically, through the
formula (\ref{F0int}). Because of the limited region in
$(\beta_0,\lambda)$ space where Cuddeford models are non-negative,
this leaves us with four models with analytical distribution
functions, corresponding to $\beta_0=\tfrac{1}{2}$, $0$,
$-\tfrac{1}{2}$ and $-1$. The most simple of them is the case
$\beta_0=\tfrac{1}{2}$, for which we obtain
\begin{equation}
    F({\cal{E}},L)
    =
    \frac{1}{4\pi^3}\,\frac{1}{L}\,
    \dfrac{3Q^2+\lambda\,(3Q^2-5Q+2)}
    {\sqrt{1+\lambda\left(\frac{1-Q}{Q}\right)^2}}.
\label{ahaha}
\end{equation}
For this distribution function, it is straightforward to check
that it remains positive for $0\leq\lambda\leq24$, in agreement
with the numerical result in Table \ref{lambdamax.tab}. For
$\beta_0=0$, we recover the Osipkov-Merritt model,
\begin{multline}
    F({\cal{E}},L)
    =
    \frac{1}{8\sqrt{2}\pi^3}
    \left\{
    \frac{3\arcsin\sqrt{Q}}{(1-Q)^{5/2}}
    \right.
    \\
    \left.
    +
    \sqrt{Q}\,(1-2Q)
    \left[\frac{8Q^2-8Q-3}{(1-Q)^2}+8\lambda\right]
    \right\},
\end{multline}
in agreement with Hernquist (1990). For the two other cases,
$\beta_0=-\tfrac{1}{2}$ and $\beta_0=-1$, the distribution
function can also be written in terms of elementary functions, but
the expressions are somewhat more elaborate.

In Fig.\ \ref{MS2799f4.eps} we show the distribution function of
the Cuddeford type for four different models. The models on the
top row have a radial central anisotropy, whereas those in the
bottom panels have a tangential anisotropy in the center. The left
and right column correspond to two different values of the
anisotropy radius. The dotted distribution functions on the
background are the distribution functions with a constant
anisotropy $\beta_0$.

The character of the Cuddeford models can directly be interpreted
from these figures. Compared to the constant anisotropy models,
the Cuddeford models have a much larger fraction of stars on
radial orbits, visible for both models with radial and tangential
central anisotropy. The most conspicuous feature of each of the
Cuddeford distribution functions is that the right part of the
$(r_-,r_+)$ diagram is completely empty, i.e.\ at large radii only
the most radial orbits are populated, which is necessary to
sustain the radial anisotropy. The boundary of the region in
turning point space beyond which no orbits are populated can be
calculated by translating the equation $Q=0$ in terms of the
turning points $r_-$ and $r_+$.
\begin{equation}
    \frac{r_-\,\psi(r_-)}{1+\lambda\,r_-^2}
    =
    \frac{r_+\,\psi(r_+)}{1+\lambda\,r_+^2}.
\end{equation}
When we substitute the Hernquist potential (\ref{hernpot}), we can
actually calculate the range of allowed orbits,
\begin{gather}
    0\leq r_-\leq r_{c,\text{max}} \\
    r_-\leq r_+\leq
    \frac{(1+r_-)+\sqrt{1+2r_-+r_-^2+4\lambda r_-^3}}{2\lambda
    r_-^2},
\end{gather}
where $r_{c,\text{max}}$ represents the radius of the largest
allowed circular orbit for a given $\lambda$,
\begin{equation}
    r_{c,\text{max}}
    =
    \frac{3^{1/3}\,(9\sqrt{\lambda}+\sqrt{81\lambda-3})^{2/3}}
    {3^{2/3}\,\sqrt{\lambda}\,(9\sqrt{\lambda}+\sqrt{81\lambda-3})^{1/3}}.
\end{equation}
Obviously, the larger $\lambda$, the more restricted the range of
allowed orbits, because the transition to radial anisotropy occurs
at smaller radii for large values of $\lambda$. This can be seen
when comparing the left and right panels of Fig.\
\ref{MS2799f4.eps}.

\subsubsection{The velocity dispersions}

\begin{figure*}
\centering
\includegraphics[clip]{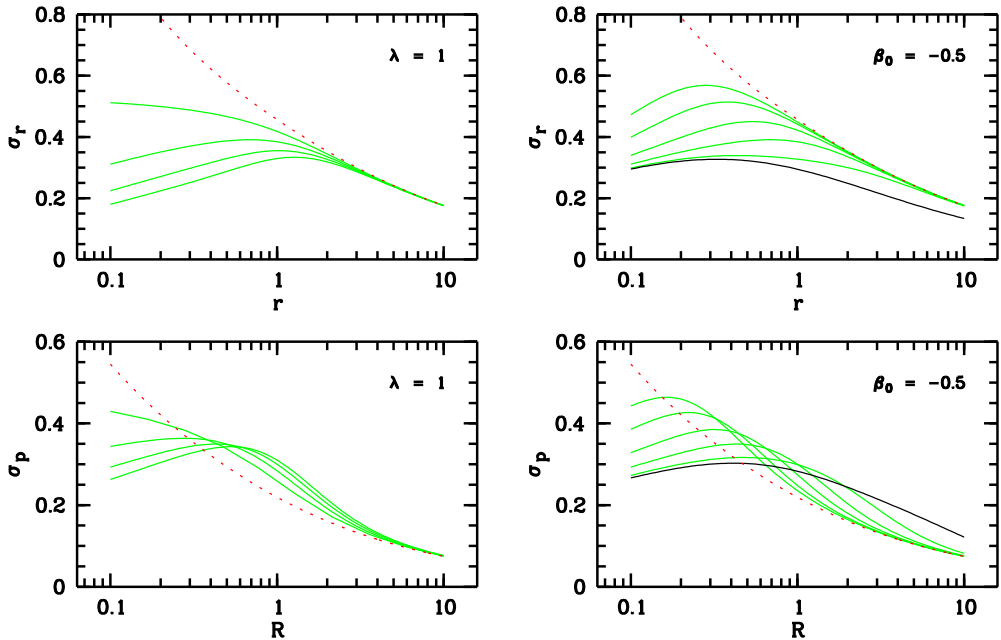}
\caption{The radial (upper panels) and line-of-sight (lower
panels) velocity dispersion profiles of the Hernquist-Cuddeford
models. The different curves in the two left panels correspond to
models with the same anisotropy radius $\lambda=1$ (i.e.\
$r_a=1$), but with a different central anisotropy parameter
$\beta_0$: plotted are $\beta=\tfrac{1}{2}$, $0$, $-\tfrac{1}{2}$
and $-1$. The dotted curves are the dispersion profiles of the
(hypothetical) completely radial Hernquist model, which
corresponds to $\beta_0=1$. The two panels on the right-hand side
contain the dispersion profiles of models with the same central,
slightly tangential, anisotropy parameter $\beta_0=-\tfrac{1}{2}$,
but with a varying anisotropy radius. The various curves
correspond to $\lambda=0$ (black solid line), 0.2, 1, 3, 8 and
$\lambda_{\text{max}}\equiv16.23$. Again, the dotted curves are
the dispersion profiles of the (hypothetical) completely radial
Hernquist model, which corresponds to $\lambda\rightarrow\infty$.}
\label{MS2799f5.eps}
\end{figure*}

In order to calculate the radial velocity dispersion associated
with models of the Cuddeford type, we use the general formula
(\ref{gensigr:def}). After some manipulation, we obtain
\begin{equation}
    \sigma_r^2(r)
    =
    \frac{r^{1-2\beta_0}(1+r)^3}{(1+\lambda r^2)^{1-\beta_0}}
    \int_r^\infty
    \frac{(1+\lambda r'{}^2)^{1-\beta_0}\txd
    r'}{r'{}^{1-2\beta_0}(1+r')^5}.
\label{cuddsigint}
\end{equation}
In general, this integral needs to be evaluated numerically, but
for the four models with integer and half-integer values of
$\beta_0$, it can be performed analytically. For example, for the
Osipkov-Merritt model $\beta_0=0$, we find
\begin{multline}
    \sigma_r^2(r)
    =
    \frac{r\,(1+r)^3}{1+\lambda r^2}
    \ln\left(\frac{1+r}{r}\right)
    \\
    -
    \frac{r\,(25+52r+42r^2+12r^3)-\lambda\,r\,(1+4r)}
    {12\,(1+r)\,(1+\lambda r^2)},
\end{multline}
which reduces to the isotropic dispersion (\ref{disp_iso}) for
$\lambda=0$. For the other integer and half-integer values of
$\beta_0$, the radial dispersion can also be expressed in terms of
algebraic functions and logarithms, but the expressions are
somewhat more elaborate.

In the top panels of Fig.\ \ref{MS2799f5.eps} we plot the radial
velocity dispersion profiles for Hernquist-Cuddeford models, for
varying $\beta_0$ and varying $\lambda$ (left and right panels
respectively). The behavior of $\sigma_r$ as a function of
$\beta_0$ is predictable. At small radii, the different models
have a different behavior, with the largest dispersion for the
most centrally radial models. At large radii they all have a
similar, purely radial, orbital structure, and as a consequence
their dispersion profiles all converge towards a single profile.
This limiting profile is the radial velocity dispersion profile
that corresponds to the (hypothetical) model with a completely
radial orbital structure, which we can obtain by either setting
$\beta=1$ in the expression (\ref{canisig}), or setting
$\beta_0=1$ in the expression (\ref{cuddsigint}),
\begin{equation}
    \sigma_r^2(r)
    =
    \frac{1}{12}\,
    \frac{1+4r}{r\,(1+r)}.
\label{sigpurerad}
\end{equation}
For a fixed central anisotropy, the behavior of the radial
dispersion as a function of the anisotropy radius also follows a
simple trend: the $\sigma_r$ profiles increase with increasing
$\lambda$, and the curves are all bounded by two limiting
profiles: on the one hand the dispersion profile (\ref{canisig})
of the constant anisotropy model (obtained by setting
$\lambda=0$), and on the other hand the hypothetical dispersion
profile (\ref{sigpurerad}) of the purely radial model (which
corresponds to $\lambda\rightarrow\infty$). Dispersion profiles
with large $\lambda$ will more quickly lean towards the purely
radial profile than models with small $\lambda$, because the
transition to a strongly radial anisotropy occurs at $r\sim
r_a=1/\sqrt{\lambda}$.

The bottom panels of Fig.\ \ref{MS2799f5.eps} show the
line-of-sight velocity dispersion of the Hernquist-Cuddeford
models. These profiles had to be calculated numerically. The
dependence of the line-of-sight dispersion upon $\lambda$ and
$\beta_0$ can be easily interpreted. In particular, the
line-of-sight dispersion profiles of the Cuddeford models tend
towards the line-of-sight dispersion profile of the hypothetical
purely radial Hernquist model, which reads
\begin{multline}
    I(R)\,\sigma_p^2(R)
    =
    \frac{R}{8}+\frac{1}{96R}
    -\frac{1}{48\pi\,(1-R^2)^2}
    \\
    \times
    \Bigl[R^2\,(20-29R^2+12R^4)\,X(R)
    +
    (2-7R^2+4R^4)\Bigr].
\end{multline}

\section{Models with decreasing anisotropy}
\label{vani.sec}

\subsection{Background}

In order to construct dynamical models with a decreasing
anisotropy, i.e.\ with a tangentially anisotropic halo, no special
inversion techniques exist, such that we have to rely on the
general formulae of Dejonghe (1986) to invert the fundamental
integral equation (\ref{detdf}). A disadvantage is that these
formulae are numerically unstable. Their usefulness is therefore
actually restricted to analytical models. But this is not
straightforward: a direct application of the inversion formulae to
an arbitrary analytical augmented density $\tilde{\rho}(\psi,r)$,
even if its looks rather simple, can result in a cumbersome
mathematical exercise, because the inversion formulae are quite
elaborate.

A useful strategy to construct models with a tangential halo
without the need to cope with the complicated general formulae, is
to profit from the linearity of the integral equation
(\ref{detdf}). In particular, it is very interesting to generate
augmented densities $\tilde{\rho}(\psi,r)$, which can be expanded
in a series of functions $\tilde{\rho}_k(\psi,r)$, which depend on
$r$ only through a power law,
\begin{equation}
    \tilde{\rho}(\psi,r)
    =
    \sum_k
    \tilde{\rho}_k(\psi,r)
    =
    \sum_k
    f_k(\psi)\,r^{-2\beta_k}.
\label{sjakamaka}
\end{equation}
Each of the augmented densities $\tilde{\rho}_k(\psi,r)$
corresponds to a dynamical model with a constant anisotropy
$\beta_k$. Combining the linearity of the integral equation
(\ref{detdf}) with the results of Sec.\ \ref{canitheory.sec}, we
find that the distribution function corresponding to the density
(\ref{sjakamaka}) reads
\begin{subequations}
\begin{equation}
    F({\cal{E}},L)
    =
    \sum_k
    F_k({\cal{E}},L),
\label{calcFk0}
\end{equation}
with
\begin{multline}
    F_k({\cal{E}},L)
    =
    \frac{2^{\beta_k}}{(2\pi)^{3/2}}\,
    \frac{L^{-2\beta_k}}{\Gamma(1-\beta_k)\,\Gamma(\tfrac{1}{2}+\beta_k)}\,
    \\
    \times
    \frac{\txd}{\txd {\cal{E}}}
    \int_0^{\cal{E}}
    \frac{\txd f_k}{\txd\psi}\,\frac{\txd\psi}{({\cal{E}}-\psi)^{1/2-\beta_k}}.
\label{calcFk}
\end{multline}
\end{subequations}
Equivalently, the moments of the distribution function can be
derived from the series expansion.

\subsection{Hernquist models with decreasing anisotropy}

\subsubsection{Construction of the density function}

For every potential $\psi(r)$, we can create an infinite number of
functions $Z(\psi,r)$ which satisfy the identity
$Z(\psi(r),r)\equiv1$. For the Hernquist potential, we can easily
create such a one-parameter family of functions $Z_n(\psi,r)$,
\begin{equation}
    Z_n(\psi,r)
    =
    \left[\psi\,(1+r)\right]^n
    \equiv
    1,
\end{equation}
with $n$ a natural number. If we multiply this family with the
density function (\ref{ell_cani}) of the constant anisotropy
Hernquist models, we create a new two-parameter family of
dynamical models, that will self-consistently generate the
Hernquist potential-density pair,
\begin{equation}
    \tilde{\rho}(\psi,r)
    =
    \frac{1}{2\pi}\,
    \frac{\psi^{4-2\beta_0+n}}{(1-\psi)^{1-2\beta_0}}\,
    \frac{(1+r)^n}{r^{2\beta_0}}.
\label{ell_vani44}
\end{equation}
Defining a new parameter $\beta_\infty = \beta_0-\tfrac{n}{2}$, we
can write this augmented density also as
\begin{equation}
    \tilde{\rho}(\psi,r)
    =
    \frac{1}{2\pi}\,
    \frac{\psi^{4-2\beta_{\infty}}}{(1-\psi)^{1-2\beta_0}}\,
    \frac{(1+r)^{2(\beta_0-\beta_{\infty})}}{r^{2\beta_0}}.
\label{ell_vani}
\end{equation}
Because we assumed that $n$ is a natural number, we can expand the
binomial in the nominator of the density (\ref{ell_vani44}), and
write it in the form (\ref{sjakamaka}), with
\begin{subequations}
\begin{gather}
    f_k(\psi)
    =
    \frac{1}{2\pi}\,
    \binom{n}{k}\,
    \frac{\psi^{4-2\beta_\infty}}{(1-\psi)^{1-2\beta_0}}
\label{vanirho0}
    \\
    \beta_k
    =
    \beta_0-\frac{k}{2},
\end{gather}
\end{subequations}
with $0\leq k\leq n$. The reason why we chose $\beta_0$ and
$\beta_\infty$ as parameters becomes clear when we look at the
expression for the anisotropy corresponding to this family of
density functions -- for the moment being without bothering
whether the density corresponds to a physically acceptable
distribution function. By means of the formula (\ref{betasplits}),
we obtain
\begin{equation}
    \beta(r)
    =
    \frac{\beta_0+\beta_\infty\,r}{1+r}.
\end{equation}
The anisotropy equals $\beta_0$ in the center and increases to
$\beta_\infty$ at large radii. Because $n$ can in principle assume
any natural number, this family of density functions hence
corresponds to dynamical models which can grow arbitrarily
tangential in the outer regions. In particular, by setting $n=0$
we recover the models with constant anisotropy
$\beta=\beta_0=\beta_\infty$ from Sec.\ \ref{cani.sec}.

\subsubsection{The distribution function}

\begin{figure}
\centering
\includegraphics[clip]{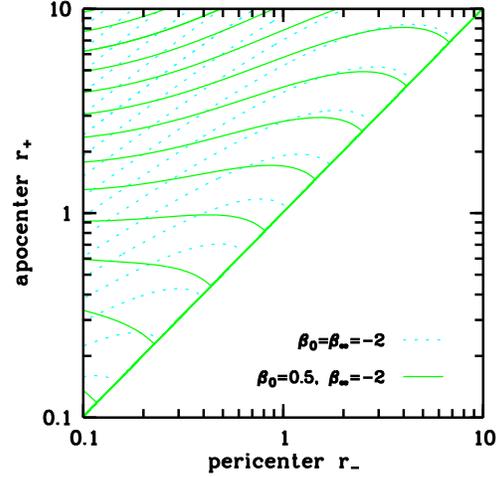}
\caption{Comparison of the distribution function corresponding to
Hernquist model with increasing anisotropy with
$\beta_0=\tfrac{1}{2}$ and $\beta_\infty=-2$, and the constant
anisotropy model with $\beta=-2$. The distribution functions are
represented as isoprobability contours in turning point space.}
\label{MS2799f6.eps}
\end{figure}

We can calculate the distribution function of these models by
applying the recipe (\ref{calcFk}) to each of the components
(\ref{vanirho0}b). We obtain after some algebra
\begin{multline}
    F({\cal{E}},L)
    =
    \frac{2^{\beta_0}}{(2\pi)^{5/2}}\,
    \Gamma(5-2\beta_\infty)\,
    L^{-2\beta_0}\,{\cal{E}}^{5/2-2\beta_\infty+\beta_0}
    \\
    \times
    \sum_{k=0}^n\,
    \binom{n}{k}\,
    \frac{1}
    {\Gamma(\tfrac{2+k}{2}-\beta_0)\,
    \Gamma(\tfrac{7}{2}-\tfrac{k}{2}-2\beta_\infty+\beta_0)}\,
    \left(\frac{L}{\sqrt{2{\cal{E}}}}\right)^k
    \\
    \times
    {}_2F_1\left(5-2\beta_\infty,1-2\beta_0;\frac{7}{2}-\frac{k}{2}-2\beta_\infty+\beta_0;{\cal{E}}\right).
\label{vanidf}
\end{multline}
This family of models is restricted by the condition
$\beta_0\leq\tfrac{1}{2}$, because for higher values of $\beta_0$
the distribution function becomes negative. For all half-integer
and integer values of $\beta_0$ (and therefore also of
$\beta_\infty$), this expression can be written in terms of
elementary functions, very analogous with the distribution
functions of the constant anisotropy models: the expression
contains integer and half-integer powers of ${\cal{E}}$ and
$1-{\cal{E}}$ and a factor $\arcsin\sqrt{{\cal{E}}}$. The models
characterized by $\beta_0=\tfrac{1}{2}$ are of a particular kind.
The hypergeometric functions in equation (\ref{vanidf}) disappears
for $\beta_0=\tfrac{1}{2}$, such that the distribution function
can be written as a finite power series of $\sqrt{{\cal{E}}}$ and
$L$.

An interesting characteristic of these models is revealed when we
look at the asymptotic behavior of the distribution function at
large radii, i.e. for ${\cal{E}}\rightarrow0$. The term
corresponding to $k=n$ will contribute the dominant term in the
sum (\ref{vanidf}), such that we obtain
\begin{multline}
    F({\cal{E}},L)
    \approx
    \frac{2^{\beta_\infty}}{(2\pi)^{5/2}}\,
    \frac{\Gamma(5-2\beta_\infty)}{\Gamma(1-\beta_\infty)\,\Gamma(\tfrac{7}{2}-\beta_\infty)}
    \\ \times
    L^{-2\beta_\infty}\,{\cal{E}}^{5/2-\beta_\infty}
    +
    \cdots.
\label{vanidfE0}
\end{multline}
This expansion is at first order independent of $\beta_0$, such
that all models with the same $\beta_\infty$ will have a similar
behavior at large radii. In particular, all distribution functions
corresponding to a particular $\beta_\infty$ will at large radii
behave as the Hernquist model with constant anisotropy
$\beta=\beta_\infty$. This is illustrated in Fig.\
\ref{MS2799f6.eps}, where we compare the distribution function of
a model with a radial core and a tangential halo with the constant
anisotropy model that has the corresponding tangential anisotropy.
At small radii, the difference between both distribution functions
is obvious: the former one has more stars on radial orbits,
whereas the latter prefers to populate circular-like orbits. At
large radii, however, the isoprobability contours of both models
agree very well.

Finally, notice that there is no analogue for this behavior at
small radii: not all models with a fixed $\beta_0$ will have a
similar behavior for ${\cal{E}}\rightarrow1$, i.e.\ at small
radii.

\subsubsection{The velocity dispersions}

\begin{figure}
\centering
\includegraphics[clip]{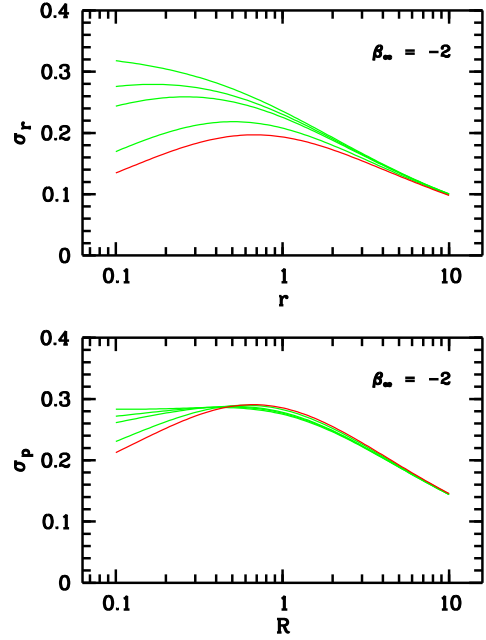}
\caption{The radial (upper panel) and line-of-sight (lower panel)
velocity dispersion profiles of Hernquist models with a decreasing
anisotropy. All models have the same tangential outer anisotropy
$\beta_\infty=-2$, but they have a different central anisotropy
parameter $\beta_0$: plotted are $\beta_0=\tfrac{1}{2}$, $0$,
$-\tfrac{1}{2}$, $-1$ and $-2$ (black line). }
\label{MS2799f7.eps}
\end{figure}

In order to calculate the velocity dispersion profiles of the
models of this type, we have various possibilities. We can either
calculate the dispersion for each of the $n$ terms
(\ref{vanirho0}) through formula (\ref{canisig:def}) and sum the
results, or directly apply the general recipe (\ref{gensigr:def}b)
on the expression (\ref{ell_vani}). In either case, we obtain an
expression very akin to the expression (\ref{canisig}) of the
models with constant anisotropy,
\begin{equation}
    \sigma_r^2(r)
    =
    r^{1-2\beta_0}\,(1+r)^{3+2\beta_\infty-2\beta_0}\,
    \text{B}_{\frac{1}{1+r}}\Bigl(5-2\beta_\infty,2\beta_0\Bigr).
\end{equation}
This expression can be written in terms of elementary functions
for all $\beta_0$ with $4\beta_0$ an integer (and hence also
$4\beta_\infty$ an integer).

Not as a surprise, the asymptotic expressions for $\sigma_r^2(r)$
for $r\gg1$ read
\begin{equation}
    \sigma_r^2(r)
    \approx
    \frac{1}{5-2\beta_\infty}\,\frac{1}{r}
    +\cdots
\end{equation}
i.e.\ they are similar to the corresponding expansions of the
constant anisotropy models with $\beta=\beta_\infty$. This
behavior is illustrated in the upper panel of Fig.\
\ref{MS2799f7.eps}, where we plot the radial velocity dispersion
profile for a set of models with varying $\beta_0$ and a fixed
$\beta_\infty$. At small radii, the models have different profiles
(those with the most radial anisotropy have the largest values of
$\sigma_r$), but at large radii, they all converge towards a
common asymptotic expansion.

The calculation of the line-of-sight velocity dispersion is also
similar to the case of constant anisotropy. It is found that
$\sigma_p(R)$ can be written in terms of elementary functions for
all integer and half-integer values of $\beta_0$, and that the
asymptotic behavior for $R\gg1$ reads
\begin{equation}
    \sigma_p^2(R)
    \approx
    \frac{8}{15\pi}\left(\frac{5-4\beta_\infty}{5-2\beta_\infty}\right)\frac{1}{R}
    +
    \cdots,
\end{equation}
which is at first order independent of $\beta_0$. An illustration
is shown in the bottom panel of Fig.\ \ref{MS2799f7.eps}.

\section{Conclusions}

Three new families of anisotropic dynamical models have been
presented that self-consistently generate the Hernquist
potential-density pair. For all models, in particular for the
Cuddeford models of Section~{\ref{cudd.sec}}, we checked the
conditions on the adopted parameters such that the distribution is
positive, and hence physically acceptable, in phase space.

They host a wide variety of orbital structures: in general, the
models presented can have an arbitrary central anisotropy, and a
outer halo with the same anisotropy, a purely radial orbital
structure, or an arbitrary, but more tangential, anisotropy. In
order to produce models that have an arbitrary anisotropy in the
central regions, and a more radial, but not purely radial,
anisotropy at large radii, the most cost-effective way seems to
construct a linear combination of a number of 'component'
dynamical models, such as the ones presented here. This technique
has been adopted for several years in the QP formalism (Dejonghe
1989, for an overview see Dejonghe et al. 2001), where most of the
components in the program libraries have an intrinsically
tangential orbital structure.

For all of the presented models, we have analytical expressions
for the distribution function and the velocity dispersions in
terms of elementary functions. They are hence ideal tools for a
wide range of applications, for example to generate the initial
conditions for $N$-body or Monte Carlo simulations. At this point,
a number of remarks are appropriate.

First, very few elliptical galaxies are perfectly spherical;
actually, various observational and theoretical evidence suggests
that many elliptical galaxies are at least moderately triaxial
(Dubinski \& Carlberg 1991; Hernquist 1993; Tremblay \& Merritt
1995; Bak \& Statler 2000). Unfortunately, an extension of the
presented techniques to construct analytical axisymmetric or
triaxial systems is not obvious, because the internal dynamics of
such stellar systems is much more complicated than in the
spherical case. Nevertheless, our models can be used as a onset to
construct numerical axisymmetric of triaxial distribution
functions with different internal dynamical structures, for
example by the adiabatic squeezing technique presented by
Holley-Bockelmann et al. (2001).

Second, the models presented here are self-consistent models,
whereas it is nowadays believed that most elliptical galaxies
contain dark matter, either in the form of a central black hole
(Merritt \& Ferrarese 2001 and references therein) and/or a dark
halo (Kronawitter et al.\ 2000; Magorrian \& Ballantyne 2001).
When constructing dynamical models with dark matter, an extra
component must be added to the gravitational potential. For
example, Ciotti (1996) constructed analytical two-component models
in which both the stellar and dark matter components have a
Hernquist density profile and an Osipkov-Merritt type distribution
function. The models presented in this paper can also be extended
to contain a dark halo or a central black hole. Indeed, the
adopted inversion techniques are perfectly suitable for this,
because the augmented density functions $\tilde{\rho}(\psi,r)$ do
not necessarily need to satisfy the self-consistency condition
(\ref{condrho}). Adding an extra term to the potential does not
conceptually change the character of the inversion, but it might
complicate the mathematical exercise.

Third, we have not discussed stability issues for the presented
models. The study of the stability of anisotropic stellar systems
is difficult, and a satisfactory criterion can not easily be
given. For stability against radial perturbations, we can apply
the sufficient criterions of Antonov (1962) or Dor\'emus \& Feix
(1973), but numerical simulations have shown that these criteria
are rather crude (Dejonghe \& Merritt 1988; Meza \& Zamorano
1997). Moreover, the only instability that is thought to be
effective in realistic galaxies is the so-called radial orbit
instability, an instability that drives galaxies with a large
number of radial orbits to forming a bar (H\'enon 1973; Palmer \&
Papaloizou 1987; Cincotta, Nunez \& Muzzio 1996). The behavior of
galaxy models against perturbations of this kind can only be
tested with detailed $N$-body simulations or numerical linear
stability analysis. Meza \& Zamorano (1997) used $N$-body
simulations to investigate the radial orbit instability for a
number of spherical models of the Osipkov-Merritt type, including
the Hernquist model. They found that the models are unstable for
$r_a\lesssim1$, which significantly restricts the set of models
that correspond to positive distribution functions (see Table
\ref{lambdamax.tab}). It would be interesting to extend this
investigation to the three families of Hernquist models presented
in this paper, but this falls beyond the scope of this paper.

\begin{acknowledgements}
The authors are grateful to Andr\'es Meza for a careful check on
the formulae derived in this paper. Fortran codes to evaluate the
internal and projected dynamics of the presented models are
available from the authors.
\end{acknowledgements}

\end{document}